\documentstyle[12pt,epsf]{article}
\renewcommand{\bar}[1]{\overline{#1}}

\def\gappeq{\mathrel{\rlap {\raise.5ex\hbox{$>$}}
{\lower.5ex\hbox{$\sim$}}}}

\def\lappeq{\mathrel{\rlap{\raise.5ex\hbox{$<$}}
{\lower.5ex\hbox{$\sim$}}}}

\begin{document}

\begin{flushright}
 CPT-2003/P.4535\\
 CERN-TH/2003-110
\end{flushright}
\bigskip\bigskip

\centerline{
\bf Positivity contraints on initial spin observables in inclusive reactions}

\vspace{22pt} \centerline{\bf  Jacques Soffer\footnote{e-mail:
Jacques.Soffer@cpt.univ-mrs.fr}}

\vspace{2pt}

{\centerline {Centre de Physique Th$\acute{\rm{e}}$orique,
CNRS, Luminy Case 907,}}

{\centerline { F-13288 Marseille Cedex 9, France, and}}

{\centerline { Theoretical Physics Division, CERN, Geneva, Switzerland}}

\vspace{4pt}
\begin{center} {\large \bf Abstract}

\end{center}

For any inclusive reaction of the type $A_1(\mbox {spin} 1/2)+ A_2(\mbox {spin} 1/2) \rightarrow B + X$,
we derive new positivity constraints on spin observables and study their implications for
theoretical models in view, in particular, of accounting for future data from the polarized $pp$ collider at BNL-RHIC.
We find that the single transverse spin asymmetry $A_N$, in the central region for several processes, for example jet production,
direct photon production, lepton-pair production, is expected to be such that $|A_N| \lappeq 1/2$, rather than the
usual bound $|A_N| \leq 1$.
\bigskip\bigskip

\centerline{PACS numbers: 13.60.Hb, 13.65.+i, 13.85.Ni, 13.88.+e}

\newpage

Let us consider an inclusive reaction of the type
\begin{equation} 
A_1(\mbox {spin} 1/2)+ A_2(\mbox {spin} 1/2) \rightarrow B + X ~,
\end{equation}
where the spins of both initial spin1/2 particles can be in any possible 
directions and no polarization is observed in the final state. The observables of this
reaction, which are the spin-dependent differential cross sections with respect to the momentum of $B$, can 
be expressed in terms of the discontinuities (with respect to the invariant mass
squared of $X$) of the amplitudes for the forward three-body scattering
\begin{equation}
A_1 + A_2 + \bar {B} \rightarrow A_1 + A_2 + \bar {B}~,
\end{equation}
as given by the generalized optical theorem. We assume parity conservation, so the complete
knowledge of this reaction requires the determination of {\it eight} real functions, which is the number
of independent spin observables \cite{GO}. 
In order to define these observables, we recall the standard notation used in Ref. \cite{BLS} $(A_1 A_2|B X)$, by
which the spin directions of $A_1,A_2,B$ and $X$ are specified in one of the three possible directions
$L,N,S$. Since the final spins are not observed, we have in fact $(A_1 A_2|0 0)$ and 
$\bf {L},\bf {N},\bf{S}$ are unit vectors, in the center-of-mass
system, along the incident momentum, along the normal to the scattering plane 
which contains $A_1,A_2$ and $B$, and along $\bf{N}\times
\bf {L}$, respectively. In addition to the unpolarized cross section $\sigma_0=(00|00)$, there
are {\it seven} spin dependent observables, {\it two} single transverse spin asymmetries
\begin{equation}
A_{1N} = (N0|00)~~~ \mbox{and} ~~~A_{2N}= (0N|00)~,
\end{equation}
and {\it five} double-spin asymmetries
\begin{eqnarray}
A_{LL}&=&(LL|00)~,~ A_{SS} = (SS|00)~,~ A_{NN} = (NN|00)~,
\nonumber \\
&& A_{LS} = (LS|00)~~ \mbox{and} ~~ A_{SL} = (SL|00)~.
\end{eqnarray}
The state of polarization of the two spin1/2 particles $A_1$ and $A_2$ is characterized by the $2 \times 2$ 
density matrices $\rho_1$ and $\rho_2$ defined as
\begin{equation}
\rho_{i}=1/2({ 1 \kern-4.6pt 1}_2  + {\bf {e}_{i}} \cdot {\bf \sigma})~~~~~  \mbox{ i = 1,2} ~~~,
\end{equation}
where $\bf {e}_1$ and $\bf {e}_2$ are the polarization unit vectors of $A_1$
and $A_2$, ${\bf \sigma} = (\sigma_x, \sigma_y, \sigma_z)$ stands for the three $2 \times 2$ Pauli matrices 
and ${ 1 \kern-4.6pt 1}_2$ is the $2 \times 2$ unit matrix. The state of polarization of the incoming system
in the reaction (1) is described by the $4 \times 4$ density matrix $\rho$, which is the direct product 
$\rho = \rho_1 \otimes \rho_2$.

The spin-dependent cross section corresponding to (1) is
\begin{equation}
\sigma({\bf {e}_1},{\bf {e}_2}) = \mbox{Tr}( M \rho)~,
\end{equation}
where $M$ denotes the $4 \times 4$ cross section matrix
which we shall parametrize in the following way \footnote{A much simpler form was used in the case
of the $pp$ total cross section, in pure spin states, to derive positivity bounds \cite{SW}.}
\begin{eqnarray}
M &=& \sigma_0[ { 1 \kern-4.6pt 1}_4 + A_{1N} \sigma_{1z}\otimes { 1 \kern-4.6pt 1}_2+ A_{2N} { 1 \kern-4.6pt 1}_2\otimes \sigma_{2z} +
 A_{NN}\sigma_{1z}\otimes\sigma_{2z} +
\nonumber\\
&& A_{LL}\sigma_{1x}\otimes\sigma_{2x} + A_{SS}\sigma_{1y}\otimes\sigma_{2y} + A_{LS}\sigma_{1x}\otimes\sigma_{2y} +
A_{SL}\sigma_{1y}\otimes\sigma_{2x}].
\end{eqnarray}
Here ${ 1 \kern-4.6pt 1}_4$ is the $4 \times 4$ unit matrix and $\sigma_0$ stands for the spin-averaged cross section.
This expression is fully justified, since we have explicitly
\begin{eqnarray}
\sigma(\bf {e}_1,\bf {e}_2)&=& \sigma_0[1 + A_{1N} e_{1z} + A_{2N} e_{2z} + A_{NN}e_{1z} e_{2z} +
\nonumber\\
&& A_{LL}e_{1x} e_{2x} + A_{SS}e_{1y} e_{2y} + A_{LS}e_{1x} e_{2y} + A_{SL}e_{1y} e_{2x}]~.
\end{eqnarray}
The crucial point is that $M$ is a Hermitian and {\it positive} matrix and in order to derive 
the positivity conditions one should write the explicit
expression of $M$ as given by Eq. (7). Then one observes that by permuting two rows and two columns, it reduces to the simple
form $\left ( \begin {array}{c|c} M_{+} & 0 \\ \hline  0 & M_{-} \end {array} \right)$, where $M_{\pm}$ are 
$2 \times 2$ Hermitian matrices which must be positive, leading to the 
following {\it two} strongest constraints \footnote{ Similar constraints were obtained in Ref. \cite{DM} for depolarization parameters
corresponding to the spin transfer between one initial spin1/2 particle and one final spin1/2 particle. For constraints on spin
observables in nucleon--nucleon elastic scattering and in the strangeness-exchange reaction $\bar {p} p \rightarrow \bar {\Lambda}
\Lambda $, see Refs. \cite{ BS, JMR}.} 

\begin{equation}
(1 \pm A_{NN})^2 \geq ( A_{1N} \pm A_{2N})^2 + ( A_{LL} \pm A_{SS})^2 + ( A_{LS} \pm A_{SL})^2~.
\end{equation}
As special cases of Eq. (9), we have the {\it six} weaker constraints

\begin{equation}
1 \pm A_{NN} \geq | A_{1N} \pm A_{2N} |~,
\end{equation} 

\begin{equation}
1 \pm A_{NN} \geq | A_{LL} \pm A_{SS} |~,
\end{equation} 
and 
\begin{equation}
1 \pm A_{NN} \geq | A_{LS} \pm A_{SL} |~.
\end{equation} 

These constraints are very general \footnote{ Let us consider three spin asymmetries whose values lie
between -1 and +1. For a simultaneous measurement of these three spin observables, the allowed region 
in a three-dimensional plot is a cube of volume $2^3=8$. However
it can be shown that inequalities like Eqs. (10,11,12) reduce strongly the allowed region, to a three-
dimensional polygon of volume 8/3. I thank J.M. Richard for this interesting observation.} 
and must hold in any kinematical region and for many different situations
such as electron--proton scattering, electron--positron scattering or quark--quark scattering, but we now turn
to a specific case, which is of direct relevance to the spin programme at the BNL-RHIC polarized $pp$ collider \cite{BSSV}.
Now let us consider a proton--proton collision and let us call $y$ the rapidity of the outgoing particle $B$. In this case 
since the initial particles are identical, we have $A_{1N}(y)=A_{2N}(-y)$ and
$A_{LS}(y)=A_{SL}(-y)$ \footnote{ I thank J.C. Collins for drawing my attention to this point and J.M. Virey for
a clarifying discussion.}.
In this case Eq. (9), which becomes two constraints among five independent spin observables, reads
\begin{eqnarray}
(1 \pm A_{NN}(y))^2 \geq ( A_{1N}(y) \pm A_{1N}(-y))^2 + ( A_{LL}(y) \pm A_{SS}(y))^2 \\
\nonumber
+ ( A_{LS}(y) \pm A_{LS}(-y))^2~.
\end{eqnarray}
This implies in particular, for $y=0$,
\begin{equation}
1 + A_{NN}(0) \geq 2| A_{N}(0) |~,
\end{equation} 
and
\begin{equation}
1 + A_{NN}(0) \geq 2| A_{SL}(0) |~,
\end{equation}
so that, the allowed range of $A_N$ and $A_{SL}$ is strongly reduced,
if $A_{NN}$ turns out to be large and negative. Conversely
if $A_{NN} \simeq 1$, these constraints are useless. 
Note that, in the kinematical region accessible to the $pp$ polarized collider,
a calculation of $A_{NN}$ for direct photon production and jet 
production has been performed \cite {SSV}; it was found that $|A_{NN}|$ is of the order of 1 or 2\%.
Similarly, based on Ref. \cite{JS}, this double transverse spin asymmetry for lepton pair 
production was estimated to be a few per-cent \cite{MSSV}. The direct consequence of these
estimates is that $| A_{N} |$ and $| A_{SL} |$, for these processes\footnote{ It is amusing to recall
that, using a phenomenological approach for lepton-pair production, 
bounds on $| A_{N} |$ larger than 50\% were obtained in Ref. \cite{ST}, but at that time it was
not known that $A_{NN}$ is small.}, are essentially bounded by 1/2. In 
addition, from Eq. (11), there are two other non-trivial constraints: $1 \geq | A_{LL} \pm A_{SS}|$.\\

Single transverse spin asymmetries in inclusive reactions at high energies are now considered to be directly related
to the transverse momentum of the fundamental partons involved in the process. This new viewpoint, which has
been advocated to explain the existing data in semi-inclusive deep inelastic scattering \cite{SMC,hermes}, will have to be
more firmly established also by means of future data from BNL-RHIC. On the theoretical side several possible leading-twist
QCD mechanisms \cite{DS,JCC} have been proposed to generate these asymmetries in leptoproduction \cite{BHS2,ADM}, 
but also in $pp$ collisions. We believe that these new positivity constraints on spin observables for a wide class
of reactions will be of interest for model builders as well as for future measurements.\\ \\

{\bf Acknowledgments} \\
The author is grateful to Prof. Tai Tsun Wu for carefully reading the
manuscript and very constructive suggestions. He also thanks X. Artru, C. Bourrely, J. Ralston and O. Teryaev
for interesting comments.

\end{document}